\documentstyle[prb,aps,epsf]{revtex}

\begin{document}
\draft

\preprint{UdeM.\ Rep.\ No.\ PMC/LJL/97-08}

\twocolumn[\hsize\textwidth\columnwidth\hsize\csname @twocolumnfalse\endcsname

\title
{\bf Self-diffusion of adatoms, dimers, and vacancies on Cu(100)}

\author{Ghyslain Boisvert\cite{byline1} and Laurent J. Lewis\cite{byline2}}

\address{
D{\'e}partement de Physique et Groupe de Recherche en Physique et Technologie des
Couches Minces (GCM), Universit{\'e} de Montr{\'e}al, Case Postale 6128, Succursale
Centre-Ville, Montr{\'e}al, Qu{\'e}bec, Canada H3C 3J7
}

\date{\today}

\maketitle

\begin{center}
{Submitted to Physical Review B}
\end{center}

\begin{abstract}

We use {\em ab initio} static relaxation methods and semi-empirical
molecular-dynamics simulations to investigate the energetics and dynamics of
the diffusion of adatoms, dimers, and vacancies on Cu(100). It is found that
the dynamical energy barriers for diffusion are well approximated by the
static, 0 K barriers and that prefactors do not depend sensitively on the
species undergoing diffusion. The {\em ab initio} barriers are observed to be
{\em significantly} lower when calculated within the generalized-gradient
approximation (GGA) rather than in the local-density approximation (LDA). Our
calculations predict that surface diffusion should proceed primarily {\em
via} the diffusion of vacancies. Adatoms are found to migrate most easily
{\em via} a jump mechanism. This is the case, also, of dimers, even though
the corresponding barrier is slightly larger than it is for adatoms. We
observe, further, that dimers diffuse more readily than they can dissociate.
Our results are discussed in the context of recent submonolayer growth
experiments of Cu(100).

\end{abstract}


\vskip2pc
]

\narrowtext

\section{Introduction}\label{intro}

Over the years, thin-film growth techniques have become pivotal in the
development of new materials and devices. In spite of this, it remains
extremely difficult, even impossible, to predict the morphology of the films
that would result from a particular set of experimental conditions ---
temperature, pressure, deposition flux, etc.\ --- even for simple
homoepitaxial systems. This state of affairs is due in large part to the fact
that growth is determined mainly by kinetic, rather than equilibrium,
effects. Since the kinetics of surfaces is determined principally by the
diffusion of atoms, either isolated (adatoms) or grouped in small clusters
(dimers, trimers, etc.), it is of utmost importance to understand in detail
the diffusion mechanisms that are involved in a given temperature range and
the rate at which they will proceed. The required information is contained in
the temperature-dependent diffusion coefficient and this is the quantity we
focus on.

By definition, the diffusion coefficient, $D$, is given by the Einstein
relation
   \begin{equation}
      D=\lim_{t \rightarrow\infty} \frac{\langle R(t)^2 \rangle}{2dt},
   \label{einstein}
   \end{equation}
where $\langle R(t)^2 \rangle$ is the mean-square displacement of the
particle undergoing diffusion and $d$ is the dimensionality of the space in
which the process is taking place. The diffusion coefficient is also often
expressed in the Arrhenius form
   \begin{equation}
      D=D_0 \exp \left( \frac{-E_A}{k_BT} \right),
   \label{arrhenius}
   \end{equation}
where $D_0$ is a ``prefactor'', $k_B$ the Boltzmann constant, $T$ the
absolute temperature, and $E_A$ the activation energy or barrier. This form
is rigorously valid in the limit $E_A \gg k_BT$.\cite{alanissila}

Experimentally, it is possible to measure directly the diffusion constant by
following the displacement of adatoms (or small clusters) over time using
field-ion microscopy.\cite{kellogg} However, because of the high imaging
field required, this technique is limited to a few materials, namely W, Ir,
Ni, Rh, and Pt. Indirect measurements are also possible, whereby the
saturation island density is measured using, for instance, scanning-tunneling
microscopy (STM) or low-energy electron diffraction (LEED), then related to
the diffusion coefficient using rate equations.\cite{venables} The problem
with such an approach lies, precisely, in the relation between island density
and diffusion coefficient (the ``scaling relation''): While it is clearly
defined when only single adatoms are mobile, it has been shown to become very
complicated when larger clusters are involved,\cite{ratsch,bartelt1} thus
making it extremely difficult to assess their relative contribution.

In this context, it becomes important to augment the experimental
measurements with detailed, accurate, calculations of the diffusion constants
based on realistic structural models. This is the route that we follow here.
More specifically, we use state-of-the-art simulation methods to calculate
the diffusion coefficients of Cu adatoms and dimers, as well as vacancies, on
the Cu(100) surface. Two distinct computational approaches are used: First,
molecular-dynamics (MD) simulations, based on the semi-empirical
embedded-atom method (EAM), are carried out; this provides us with
qualitative knowledge of the processes involved during diffusion, as well as
quantitative information on the diffusion coefficients for various
mechanisms, in particular the prefactors. This is extremely important since
only dynamical simulations can provide accurate values for the prefactors,
which can vary significantly as a function of the barrier
height.\cite{boisvert1} The activation energies, in contrast, can be
calculated accurately using static (0 K), energy-minimization methods, as we
will demonstrate. In order to go beyond the approximate EAM, we have
performed, second, detailed {\em ab initio} calculations of the energy
barriers. Calculations were done within the framework of density-functional
theory (DFT),\cite{kohn} using both the all-electron (AE), full-potential,
linear-muffin-tin-orbital method (FP-LMTO),\cite{meth1} and the now-standard
pseudopotential-plane-wave (PP-PW) approach.\cite{payne} In the latter case,
since it has been shown that the inclusion of gradient corrections to the
exchange-correlation energy leads to a significant improvement over the usual
local-density approximation (LDA) for $3d$ metals,\cite{khein,philipsen} most
calculations have been performed in the generalized-gradient approximation
(GGA); some LDA results are nevertheless presented in order to compare with a
previous study by Lee and coworkers.\cite{lee}

The reasons for studying the system Cu/Cu(100) are manifold. First,
simplicity: the surface lattice is square and, because the system is
homoepitaxial, only one type of chemical species need be considered (i.e.,
modeled) and no large stress, e.g., arising from mismatch, will be involved.
Such simple systems --- homoepitaxial face-centered cubic metals --- have
been the object of numerous studies of the fundamental aspects of growth (see
for instance Refs.\
\onlinecite{bartelt1,ernst1,ernst2,zuo,bartelt2,breeman0,barkema,biham,bartelt3,bartelt4}).
Second, surface diffusion on Cu(100) has been studied in detail both
experimentally\cite{miguel,breeman1,ernst3,durr} and theoretically.
\cite{lee,liu1,hansen1,sanders,hansen2,breeman2,perkins1,black,liu2,stolze,perkins2,karimi,merikoski,evangelakis,shi,kumar}
(Of the latter, only Ref.\ \onlinecite{lee} is a first-principles
calculation.) Yet, no clear picture has emerged: While experiment indicates a
barrier in the range 0.28 to 0.40 eV, but gives no information on the actual
mechanism {\em via} which diffusion takes
place,\cite{miguel,breeman1,ernst3,durr} the {\em ab-initio} calculations of
Lee {\em et al.}\cite{lee} predict that diffusion proceeds primarily through
simple hopping of an adatom on the surface, with a barrier height of 0.69 eV.
In comparison, studies based on various semi-empirical potentials give values
in the range 0.20--0.70 eV and are in disagreement on the preferred diffusion
mechanism.
\cite{liu1,hansen1,sanders,hansen2,breeman2,perkins1,black,liu2,stolze,perkins2,karimi,merikoski,evangelakis,shi,kumar}

\begin{figure}
\vspace*{-0.25in}
\epsfxsize=3.5in \epsfbox{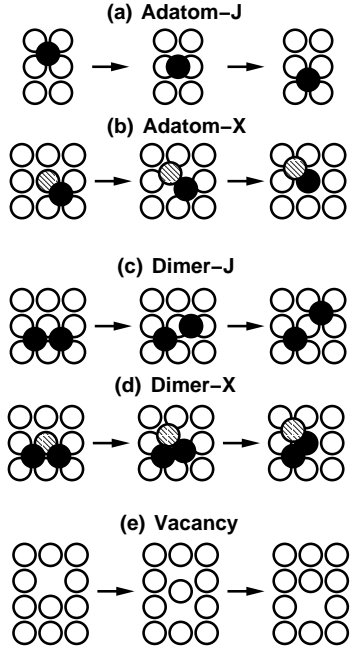}
\vspace{1.5in}
\caption{
The various diffusion processes studied in the present work; J and X refer
to jump and exchange, respectively.
\label{diff_proc}
}
\end{figure}

Evidently, the dominant mechanism for diffusion on Cu(100) is not resolved.
The various processes examined here --- jump and exchange for the adatom and
the dimer, and jump only for the vacancy --- are illustrated in Fig.\
\ref{diff_proc}. Although there are other possibilities, especially at high
temperatures,\cite{black} the ones we consider are found, from our MD
simulations, to be the best candidates for low-temperature diffusion. We
computed, in addition, the diffusion coefficient for an adatom moving along a
step in order to understand the shape of islands on (100) terraces.

Our LDA calculations of the energy barrier for adatom diffusion corroborate
the previous study, also within LDA, by Lee {\em et al.};\cite{lee} however,
we find the barrier to be {\em significantly} reduced when calculated within
the GGA, thus bringing it much closer to the experimental value. In any case,
the preferred mechanism for diffusion is found to be hopping. We find the
barrier for dimer diffusion to be close to that for the adatom, but lower
than that for dimer dissociation. We also find that vacancies are more mobile
than adatoms and that diffusion of adatoms along a step proceeds much more
rapidly than on a terrace so that the island shape during growth should be
close to equilibrium. Prefactors, finally, do not depend sensitively on the
species undergoing diffusion. These findings are discussed in the context of
recent growth experiments performed on this system.
\cite{miguel,breeman1,ernst3,durr}

\section{Computational details}\label{model}

\subsection{Semi-empirical calculations}\label{eam}

As mentioned already, the atoms in the MD simulations were assumed to
interact {\em via} the EAM potential proposed by Foiles, Baskes, and
Daw;\cite{foiles} the optimized parameterization of Adams, Foiles, and
Wolfer\cite{adams} was used. Although this model has been fitted to bulk
properties, it has been applied successfully to the study of various surface
phenomena.\cite{foiles2}

For the MD calculations on the flat (100) surface, we used a geometry and a
procedure similar to our previous study of adatom diffusion on Ag and Au
surfaces.\cite{boisvert2} The surfaces were approximated by slabs containing
8 layers (excluding the adatom) of which the bottom two are held fixed in
order to mimic the bulk. Each of the layers contains 64 atoms. For the
diffusion along a step, we considered a (13,1,1) surface, which is vicinal to
the (100) surface, and possesses 6-atom wide terraces. In order to keep the
rectangular shape of the unit cell, two steps are included at the surface.
(In the (100) direction, the planes are stacked in the order $ABAB...$). Each
of the two terraces contains 36 atoms and we thus have 72 atoms per layer.
The same number of layers as in the case of the (100) surface was used. When
studying diffusion, an adatom is added at each of the two steps. In all
cases, periodic boundary conditions were applied in the lateral directions,
i.e., parallel to the surface, so that the system is effectively infinite in
the $x-y$ plane. The lattice parameter of the rigid layers was determined
from a series of runs on the bulk material in the $(N,P,T)$ ensemble (using a
256-atom system) at each simulated temperature. All others simulations are
carried out in the $(N,V,T)$ ensemble.

In most cases studied here, it is necessary to deal with more than one
diffusion mechanism for a given species, viz.\ jump and exchange. It is
simpler, then, to consider the frequency at which each type of event is
taking place, rather than the actual rate of diffusion as given by Eq.\
\ref{einstein}. As we observed in a previous article,\cite{boisvert2} when
the barriers are high enough, diffusion can be assimilated to a random walk,
so that the diffusion coefficient is simply related to the frequency of
events. We will see, in Sec.\ \ref{MD}, that the barriers are indeed high
compared to the temperature at which diffusion is considered. In order to
determine the frequencies, the evolution of the diffusing species (adatom,
dimer, or vacancy) is followed at several temperatures (see Sec.\ \ref{res})
for a time long enough to yield reliable and reproducible statistics. In
practice, the runs consisted of a period of equilibration of 48 ps, followed
by a period of ``production'' of 5--18 ns (depending on the number of events
observed, i.e., barrier and prefactor for the process, as well as
temperature), during which statistics were accumulated. A timestep of 2.4 fs
was used to perform the numerical integration of the equations of motion. To
speed up the calculations, we introduced a cutoff distance in the potential,
beyond which all interactions were neglected. It was found that a cutoff of
4.8 \AA\ (between third and fourth-neighbor shells) yields barriers within
5\% of their converged values except in the case of the dimer, for which a
cutoff of 5.4 \AA\ (between fourth and fifth-neighbor shells) was necessary
to achieve the same level of accuracy.

In order to determine the 0 K, static barriers and compare them with their
dynamical equivalents, we also carried out a series of energy-minimization,
``molecular-statics'' (MS), calculations, whereby the atoms are moved
iteratively in the directions of the forces acting on them until these vanish
(``relaxation''). The static energy barrier is obtained by relaxing the
system in both the equilibrium and the transition state; in the latter case,
a constraint is used to maintain the particle(s) at the saddle point and
minimization is carried out with respect to all other degrees of freedom.

\subsection{{\em Ab initio} calculations}\label{abinitio}

Since first-principles MD for transition metals is too demanding for a direct
study of diffusion, only MS calculations of the barriers were carried out
using this approach. First, following our study of diffusion on Ag, Au, and
Ir,\cite{boisvert3} we computed the energy barriers using the
FP-LMTO.\cite{meth1} This method is approximate only in the parameterization
of the exchange-correlation energy; however, it provides no analytical forces
on the ions so that relaxation effects cannot be estimated accurately. To
overcome this problem, we also performed, second, calculations using the
PP-PW approach, where the ionic-core potential is approximated by a
pseudopotential. In this case, fully self-consistent calculations were
carried out using both the LDA and the GGA, whereas only the LDA was used in
the case of FP-LMTO. We now describe our computational approach in more
detail.

\subsubsection{FP-LMTO}\label{fplmto}

For the FP-LMTO calculations, we used the same approach as in Ref.\
\onlinecite{boisvert3}. The surface was constructed in supercell geometry,
and consisted of a slab of 5 to 9 layers and a vacuum region of about $10$
\AA\ periodically replicated in space; each layer contains 4 to 9 atoms. Both
the number of layers and the number of atoms per layer were varied in order
to ensure convergence with respect to system size (see below). To determine
the barriers for diffusion, an adatom was placed on each of the two external
surfaces of the slab, with the central layer taken as a mirror plane in order
to reduce the computational load. Only adatom jump diffusion was considered
using this technique. The $z$ coordinate of the adatom was varied in order to
minimize the total energy of the slab. All other atoms were kept in their
ideal, bulk-like position, except for the surface layer, which was relaxed
before the adatom was introduced (i.e., in its clean state) using a 5-layer
$(1\times1)$ unit cell.

To compute the energy, we used a basis set of 27 functions per atom,
consisting of $4s$, $4p$, and $3d$ functions with kinetic energy $-\kappa^2=$
$-$0.7, $-$1.0, and $-$2.3 Ry, respectively. Scalar-relativistic corrections
were included and the exchange-correlation energy evaluated using the
Ceperley-Alder form.\cite{ceperley} The integration over the Brillouin zone
employed 36 equidistant {\bf k} points when using 4 atoms per layer, and 16
when using 9 atoms per layer; a Gaussian broadening of 20 mRy was used to
ensure the numerical stability of the integral. Bulk and clean surface
properties were calculated using the same density of {\bf k} points.

\subsubsection{PP-PW}\label{pppw}

In the PP-PW approach, the core orbitals are replaced by pseudopotentials.
Here, we used pseudopotentials generated according to the semi-relativistic
scheme of Troullier and Martins,\cite{troullier} and cast in the
fully-separable, norm-conserving form of Kleinman and Bylander, with the $s$
component only being local.\cite{kleinman,gonze,fuchs} The $3d$ electrons
were treated as valence states. The electronic wavefunctions were expanded in
plane waves with a kinetic energy cutoff of 60 Ry in the LDA\cite{ceperley}
and 65 Ry in the GGA.\cite{perdew} The {\bf k}-space integration was
performed using a set of 9 equidistant points in the surface Brillouin zone
for the systems with 4 atoms per layer and 4 points for the ones with 9 atoms
per layer. To improve convergence, the electronic states were occupied
according to a Fermi distribution with a temperature of $k_BT_{el} = 0.1$ eV
and the total energy extrapolated to zero electronic temperature. For similar
reasons, the {\em initial} wave-functions were obtained from the
self-consistent solutions of the Kohn-Sham Hamiltonian in a mixed-basis set
composed of pseudo-atomic orbitals and plane waves with kinetic energy less
than 4 Ry.\cite{kley} The minimization of the energy with respect to the
electronic degrees of freedom was done iteratively using a
Car-Parrinello-like technique.\cite{carpar,stumpf}

In view of the high energy cutoff needed in the plane-wave expansion, it is
important to keep the system size to a minimum. To do so, we used a geometry
slightly different from that described above, considering here a single
adatom on one surface of the slab. This enables us to use a smaller number of
layers and, therefore, a smaller supercell. In practice, 3 to 7 layers were
considered, with only the adatom and at most the top two layers allowed to
relax. Damped Newton dynamics was used to displace the atoms; this was done
iteratively until all forces (on the atoms allowed to relax) became less than
0.01 eV/\AA. Bulk and clean surface properties were calculated using the same
{\bf k}-point density as in the diffusion study; for the clean surface, a
9-layer, $(1\times1)$ cell was used.

\subsubsection{Bulk and clean (100) surface}\label{bulk_res}

In order to establish the validity of our {\em ab initio} approach, we have
computed, prior to considering diffusion, the bulk lattice constant and some
properties of the clean (100) surface, namely the surface energy, surface
relaxation, and work function. The results are listed in Table \ref{tests}
along with other {\em ab initio} results and available experimental data.

For the lattice constant, first, we get good overall agreement with previous
calculations. It is well known that the LDA underestimates lattice constants.
The GGA, however, tends to overcompensate and, as a result, the GGA lattice
constants are usually larger than experiment,\cite{khein} as indeed found
here. Also, as noted by Lu and coworkers,\cite{lu} pseudopotential
calculations yield lattice constants larger than all-electron (AE)
calculations, also a feature observed here. The combination AE-GGA,
therefore, seems to be optimal (but not available to us at present); this is
also supported by the fact that the GGA provides a much better description of
the cohesive energy than the LDA.\cite{philipsen} Thus, even though the
PP-PW-GGA combination does not yield accurate lattice constants, it is better
suited to describing Cu than PP-PW-LDA.

For the clean (100) surface, now, our results are also in relatively good
agreement with other calculations, when available, and with experiment. It is
interesting to note that, even if AE and PP calculations give different
values for the lattice constant, they lead to very similar surface properties
for a given level of approximation of the exchange-correlation energy (i.e.,
LDA or GGA). Thus, a self-consistent PP calculation is quite suitable to
describe surface properties here, even if bulk properties are not as well
described as in AE calculations. We note from Table \ref{tests} that the GGA
reduces the surface energy and the work function compare to the LDA. This
effect of the GGA on metallic surfaces has already been predicted from
jellium calculations.\cite{perdew} The same phenomenon has been observed on
Cu(111),\cite{boisvert5} Pt(111),\cite{boisvert4} and Ag(100).\cite{yu} In
view of the difficulty in measuring accurate surface energies, and the
scatter in the experimental values for the work function, it is not clear
which exchange-correlation functional best describes surfaces properties.
However, considering that the GGA provides a better description of bulk Cu,
we conclude that it is better suitable, also, for Cu surfaces.

\section{Results}\label{res}

As we previously have shown,\cite{boisvert2} adatom diffusion barriers can be
reliably extracted from static calculations. However, in order to determine
completely the diffusion coefficient, the prefactor is also needed, and there
seems to be no simple way of extracting this quantity with sufficient
accuracy from purely static calculations. On the other hand, first-principles
MD simulations are too demanding in terms of computer time for such an
enterprise to be undertook, and one must therefore resort to classical models
in order to calculate the prefactors. We present here the results of our
study of diffusion on Cu(100) using both a classical and a quantum
description of forces for calculating the energy barriers, and classical MD
for estimating the prefactors.

It is often assumed, for convenience and without much justification, that
prefactors for diffusion are constant (see, e.g., Ref.\
\onlinecite{kellogg}), independent on the details of the surface. However, in
a recent study,\cite{boisvert1} we have shown that the diffusion of adatoms
follows the compensation (Meyer-Neldel) law and, as a result, prefactors can
vary by several orders of magnitude. The Meyer-Neldel rule states that, for a
family of Arrhenius processes,
   \begin{equation}
   X = X_0 \exp( - E_A / k_BT )
   \label{mna}
   \end{equation}
(which is the case of diffusion, Eq.\ \ref{arrhenius}), the prefactor $X_0$
depends exponentially on the activation energy $E_A$:
   \begin{equation}
   X_0 = X_{00} \exp( E_A/\Delta_0 ),
   \label{mnr}
   \end{equation}
where $\Delta_0$ is the iso-kinetic (or Meyer-Neldel) energy and $X_{00}$ is
a constant. It is therefore important, in order to determine the most mobile
species in a given temperature regime, to see how prefactors compare. These
results will be presented in Sec. \ref{MD}. We discuss, first, the static
energy barriers, both in the context of EAM and from first principles.

\subsection{Static energy barriers -- EAM}\label{eam_ms}

\subsubsection{Adatoms}\label{eam_ad_ms}

Our results for the static barriers on Cu(100), $E_A^0$, for the various
cases of diffusion considered here, are listed in Table \ref{para}; our
results generally agree with previous estimates using a similar theoretical
framework.\cite{liu1,liu2,karimi,shi} The adatom, within the EAM picture, is
found to diffuse preferably {\em via} a jump mechanism, the barrier for
exchanges being much higher --- 0.73 vs 0.50 eV. From these values of the
barriers, one would conclude that exchange diffusion contributes negligibly
to mass transport (in comparison to jump diffusion). However, as we will see
in Sec.\ \ref{MD}, this conclusion must be taken with caution because the
prefactor for exchanges is much larger than that for jumps.

\subsubsection{Dimers} \label{eam_dim_ms}

In order to determine if small clusters are mobile at low temperatures, or
if, rather, they are more likely to dissociate, we have calculated the
barriers for the jump and exchange diffusion of dimers, as well as the
binding energy, dissociation energy, and excess energy of metastable vs
equilibrium state. The results are listed in Tables \ref{para} and \ref{dim}.

Just like adatoms, dimers diffuse much more easily by jumps than by
exchanges, and the barriers for the two processes are very similar to the
corresponding ones for adatoms. Thus, as far as the mechanism is concerned,
adatoms and dimers behave in the same way; if we consider only the barriers,
dimers are expected to be mobile at the same temperature as the adatoms.

It is important to note that, for dimers, the barrier for jump diffusion
(0.49 eV --- cf.\ Table \ref{para}), is the barrier to go from the
equilibrium to the metastable state, as depicted in Fig.\ \ref{diff_proc}(c).
Indeed, the barrier to go from the metastable to the equilibrium state, which
is equal to the barrier height minus the excess energy of the metastable
state ($0.49-0.29=0.20$ eV --- cf.\ Table \ref{dim}), is much smaller than
the reverse; the corresponding process thus occurs much faster. The limiting
process, therefore, is the one considered here, i.e., equilibrium to
metastable.

The dissociation barrier for a dimer is given, approximately, by the sum of
the diffusion barrier for the adatom and the binding energy of the
dimer.\cite{bartelt1} In the present case, this leads to a barrier of 0.85
eV. Direct calculation of the dissociation barrier is difficult considering
that there are several possible dissociation pathways. We have examined
different possibilities and found the lowest barrier to be 0.81 eV
(corresponding to a 50\% stretch of the dimer along its equilibrium axis), in
good agreement with the approximate value above, and significantly larger
than the barriers for diffusion. Thus, dimers are already mobile at
temperatures well below the onset of dissociation.

We now compare mass transport from dimers and adatoms, considering only the
predominant jump-diffusion process. To do so, it is necessary to first
determine the mean-square displacement of the center of mass of the dimer
during an event. When a dimer jumps, there exists 4 different paths leading
to a zero net displacement of the center of mass, 4 leading to a displacement
of $a^2/2$, and 8 leading to a displacement of $a^2/4$, where $a$ is the
lattice constant. On average, therefore, the mean-square displacement is
$a^2/4$. This is a factor of 2 smaller than the corresponding displacement
for an adatom, but the dimer contains 2 atoms; hence, as much mass is
transported in a single event as is in the case of adatoms, on average. We
are thus led to conclude that, within EAM, {\em dimers contribute as much to
mass transport as adatoms.}

\subsubsection{Vacancies} \label{eam_vac_ms}

While vacancy diffusion is not, {\em per se}, a mechanism for growth, it can
have important consequences on mass transport, in particular in the process
of annealing defected surfaces. In Fig.\ \ref{diff_proc}, we show the
mechanism by which a vacancy diffuses on the (100) surface --- basically a
jump. The corresponding barrier is 0.47 eV (cf.\ Table\ \ref{para}), larger
than the value of 0.35 eV reported in Ref.\ \onlinecite{karimi}, which is in
error.\cite{karimi2}

The jump-diffusion barrier for vacancies is, also, close to that for adatoms.
Of course, the actual contribution of each process depends on the relative
population of the two species, which itself depends on the formation
energies; indeed, the migration energy is the sum of the formation energy and
the diffusion barrier. Using EAM, Karimi and coworkers\cite{karimi} found
formation energies of 0.59 and 0.71 eV for the vacancy and the adatom,
respectively. Thus, vacancies have lower formation energy than adatoms and
should therefore contribute more to mass transport. However, during growth, a
large reservoir of adatoms is available, and their mobility is limited only
by the diffusion barrier. In contrast, vacancies first have to form, i.e.,
their mobility is determined by the diffusion barrier {\em plus} the
formation energy, and thus severely reduced, to the point where their
contribution to mass transport will in fact be negligible at temperatures of
interest.

\subsubsection{Steps}

The shape of islands on otherwise flat terraces is important for a proper
understanding of growth phenomena. It is determined, in equilibrium
conditions, by the energies of the various steps defining its perimeter.
During growth, the equilibrium shape can be attained only if the kinetic
processes leading to equilibrium are fast enough to overcome the continuous
arrival of new adatoms onto the island. For the (100) surface, the
equilibrium island shape is approximately square, with the corners rounded.
The sides of the island are formed by $\langle$110$\rangle$-oriented steps,
which are the most densely packed on this surface. Thus, if diffusion along
these steps, which measures the rate at which equilibrium is reached, is fast
compared to diffusion on a terrace, corresponding to the rate at which
adatoms arrive, then the shape will be close to equilibrium. We find, for
diffusion along $\langle$110$\rangle$ steps, a barrier value of 0.26 eV (cf.\
Table \ref{para}); this is indeed much lower than the barrier for adatom
diffusion on terraces, 0.50 eV. Thus, the shape of islands is expected to be
close to equilibrium {\em even} during growth.

\subsection{Static energy barriers -- {\em Ab initio}}\label{ab_ms}

\subsubsection{Adatoms}\label{ab_res_ad}

In order to assess the validity of the EAM calculations, we move on with a
discussion of {\em ab-initio} diffusion barriers, starting with the case of
adatoms which, in view of the discrepancy between the first-principles
calculations of Lee {\em et al.}\cite{lee} and experiment, constituted the
initial motivation of this work. Also we have, for adatoms, carried out an
extensive study of convergence with respect to size and other parameters of
the model; the results are presented in Table \ref{adatom}.

A first observation from Table \ref{adatom} is that, within numerical
accuracy, the AE-FP-LMTO and PP-PW calculations give the same result for the
jump-diffusion barrier for cells of equivalent size, at the same level of
approximation (compare, e.g., the FP-LMTO-LDA and PP-PW-LDA for jumps on the
$(2\times2)$ cell with 5 or 7 layers). Thus, the use of PP's seem to have
little effect on diffusion barriers even if it yields lattice constants
different from AE calculations. The same behavior was observed for clean
surface properties, as mentioned in Sec.\ \ref{bulk_res}. This establishes
the validity of the approach and only PP calculations will therefore be
discussed from now on.

Within the PP-PW scheme, forces on the ions are easy to compute and the
effect of relaxation on diffusion barriers can be assessed. This question was
neglected in our previous study of self-diffusion on Ag, Au, and Ir surfaces
using the FP-LMTO technique, for which analytical forces are not available in
the supercell geometry.\cite{boisvert3} From the PP-PW-LDA results for the
3-layer, $(2\times2)$ cell given in Table \ref{adatom}, it is clear that the
effect of relaxation on the barrier for jumps is negligible (0.75 eV for the
unrelaxed surface, indicated by the superscript ``$u$'', vs 0.74 eV for the
relaxed surface). Evidently, this is more important for the exchange process,
but nevertheless small (1.23 vs 1.18 eV, i.e., less than 5\%) --- smaller in
fact than could be expected.

In most calculations, only the top layer of the slab was allowed to relax (in
addition to the adatom). We have verified that this is not a limiting
approximation by carrying out some calculations where, also, the second layer
was relaxed. This is indicated by the superscript ``$2$'' in Table
\ref{adatom} for the 4-layer, $(2\times2)$ cell under PP-PW-LDA. The effect
is extremely small, no more than 0.01 eV, i.e, within the accuracy of the
method. One must not generalize these conclusions to other systems, however,
especially the (111) surface of fcc metals where barriers for jumps are
small. For instance, for Cu diffusion on Cu(111), the barrier drops by a
factor of almost two, from 0.14 to 0.08 eV, when allowing the first atomic
layer to relax;\cite{boisvert5} similar effects are also found for
Pt/Pt(111).\cite{boisvert4}

The convergence with respect to supercell size was examined very carefully.
As can be seen from the PP-PW-LDA results in Table \ref{adatom}, the barriers
for both processes ``oscillate'' slightly when increasing the number of
layers beyond 4. In the case of exchanges, the fluctuations in the barrier
height are more important than for jumps in absolute value, but quite similar
on a relative scale, viz.\ about 10\%. We note also that the barrier for
jumps does not change noticeably upon increasing the lateral size of a
3-layer slab from $(2\times2)$ to $(3\times3)$, while the barrier for
exchanges drops by about 11\%.

Since the barrier for jumps remains the same upon going from a $(2\times2)$
to a $(3\times3)$ cell, we conclude that our error on this energy is
essentially that arising from the convergence with respect to the number of
layers, i.e., about 10\%. For exchanges, we observe the barrier to vary a bit
upon going from a $(2\times2)$ to a $(3\times3)$ cell, but we expect that it
should not change substantially for larger systems. Thus, the error on the
barrier for exchanges is expected to be about the same as for jumps, namely
about 10\%.

Finally, we also verified the convergence of the results with respect to
Brillouin-zone sampling, again for the 4-layer, $(2\times2)$ cell under
PP-PW-LDA. We found, upon increasing the number of (surface) {\bf k}-points
from 16 to 25, the barriers for jumps and exchanges to change very little ---
by 0.02 and 0.03 eV, respectively, considering only relaxation of the top
substrate layer.

As mentioned earlier, we know of only one other {\em ab-initio} calculation
of the barriers for adatom diffusion on Cu(100), by Lee and
coworkers,\cite{lee} carried out within the LDA. Using a 3-layer $(3\times3)$
cell, they found activation energies of 0.69 eV for jumps and 0.97 eV for
exchanges. This compares quite well with our results for the same cell size,
as can be seen in Table \ref{adatom}. The small differences are likely due to
different Brillouin-zone sampling schemes: while we used a $2\times2$ grid of
equidistant points for this cell, Lee {\em et al.} employed only the $\Gamma$
point.

The GGA, as we have seen above, yields a better description of bulk Cu
properties, such as lattice constant and cohesive energy, compared to the
LDA.\cite{khein,philipsen} We have also found in Sec.\ \ref{bulk_res} that it
has an effect on surface properties such as the surface energy and the work
function. It is therefore of interest to see how diffusion barriers compare
in the two approximations. This question was addressed recently in the case
of Ag/Ag(100) by Yu and Scheffler;\cite{yu} for Ag, the GGA is known to
overcompensate the LDA error as far as the lattice constant is
concerned.\cite{khein} Yu and Scheffler found, under the GGA, the barriers to
drop from 0.52 to 0.45 eV in the case of jumps, and from 0.93 to 0.73 eV for
exchanges, a decrease of respectively 13\% and 22\% from the LDA value. In a
recent experiment, Langelaar {\em et al.}\cite{langelaar} found a diffusion
barrier of $0.43 \pm 0.02$ eV, assuming a prefactor of 10 THz as obtained
from MD simulations.\cite{boisvert2} This agrees within error with the above
GGA value for jumps; however, the LDA value is not far either and it is
therefore difficult to say which approximation is better. In the present
case, the GGA barriers are about 22\% smaller than the corresponding LDA
values (cf.\ Table \ref{adatom}). This is larger than the numerical accuracy
estimated earlier --- about 10\%. We are therefore led to conclude that the
GGA leads to a significant decrease of energy barriers compared to the LDA.

Our best estimates for the activation energies are thus $0.52 \pm 0.05$ and
$0.96 \pm 0.10$ eV, for jumps and exchanges, respectively. Thus, just as was
the case with EAM, the adatom is found to diffuse more readily {\em via} a
jump mechanism. In fact, as can be seen from Table\ \ref{para}, the EAM
barrier for jumps is in quantitative agreement with the GGA barrier. We thus
expect MD/EAM simulations to yield a reliable, {\em quantitative} estimate of
the prefactor for jump diffusion (see below). For exchanges, the EAM
underestimates the barrier with respect to the GGA and, therefore, the MD
simulations can only yield qualitative information.

\subsubsection{Dimers}\label{ab_res_dim}

For dimers, now, we have not performed detailed convergence tests, but error
bars can be estimated from the above convergence study for adatoms: Since the
coverage for a dimer on a $(3\times3)$ cell is comparable to that for a
single adatom on a $(2\times2)$ cell, the error on the barriers for exchange
diffusion should be approximately the same, namely 20\%, while other
quantities --- barriers for jumps, binding energies, and excess energies of
the metastable configuration --- should be accurate to about 10\%. All the
results discussed below refer to a 4-layer, $(3\times3)$ unit cell.

Dimers are found to diffuse preferentially {\em via} jumps, as was the case
also for adatoms, with a barrier of 0.57 eV compared to 0.79 eV for exchanges
(cf.\ Table \ref{para}). The barrier for jumps estimated from EAM compares
well with the GGA value, as was also true of adatoms, although the deviation
here is a bit larger. For exchanges, the EAM and PP-PW-GGA estimates are in
good agreement, but in view of the large error bar on the latter, it is
difficult to ascertain that this agreement is genuine.

As already noted in Sec.\ \ref{eam_dim_ms}, the barrier towards dissociation
is given, roughly, by the sum of the dimer binding energy and the diffusion
barrier of the adatom.\cite{bartelt1} No attempt to compute this quantity
directly from first principles has ever been made because of the
prohibitively large system size required. Using the approximate form, we
obtain a dissociation barrier of 0.74 eV (cf.\ Table \ref{dim}), much higher
than the barrier for diffusion. Thus, we conclude that dimers are mobile at
temperatures lower than those for which dissociation takes place. This is in
qualitative agreement with EAM, even though quantitatively, the difference
between barriers for diffusion and for dissociation is larger within EAM than
within GGA, due to the combined effect in EAM of a lower diffusion barrier
and a higher dissociation barrier. The difference in the dissociation barrier
can be traced back to the dimer binding energy, which is much lower within
GGA than within EAM.

In view of the large excess energy of the metastable configuration with
respect to equilibrium, 0.35 eV, which agrees well with EAM, the discussion
on mass transport presented in Sec.\ \ref{eam_dim_ms} remains valid: the
relative contributions to mass transport by adatoms and dimers can be
determined solely on the basis of their jump frequencies. Since the energy
barrier for dimer diffusion is slightly larger than that for adatoms --- 0.57
vs 0.52 eV --- we conclude that adatoms will be mobile at lower temperatures
than the dimers. In view of the small difference, however, the temperature
range in which the above conclusion is valid will be rather narrow.

\subsubsection{Vacancies}

The barrier for the diffusion of vacancies was also determined {\em ab
initio} using the GGA. Its value, given in Table\ \ref{para} along with other
barriers, is estimated to be 0.42 eV, with an error bar of at most 20\%.
Again, the agreement with the EAM result, 0.47 eV, is striking. Also, this is
smaller than the barrier for adatom jump diffusion. Thus, vacancy diffusion
should dominate mass transport on the surface except, as discussed in Sec.\
\ref{eam_vac_ms}, during growth, when a large ``reservoir'' of adatoms is
available.

As can be concluded from Table \ref{para}, the present EAM parameterization
provides, in most cases, a very satisfactory agreement with the
first-principles results we have just described, taking due account of the
uncertainties of the {\em ab initio} calculations. This is a bit of a
particular case, however: we have shown, in a recent
publication,\cite{boisvert3} that the agreement between EAM and
first-principles calculations could be quite acceptable when the barriers are
large, but poor when they are small, as is the case for instance on the (111)
surface of fcc metals.

\subsection{MD-EAM}\label{MD}

\subsubsection{Adatoms}

While static calculations of diffusion barriers on Cu(100) have been
numerous, direct simulations of the actual diffusion processes have been
rather scarce.\cite{breeman0,black,merikoski,evangelakis,raeker} Of these,
only Ref.\ \onlinecite{evangelakis} is concerned with a detailed Arrhenius
study of adatom diffusion, and dimer and vacancy diffusion was not
considered. Also, the model used, based on a tight-binding description of the
interatomic potentials, differs from ours. For consistency, and in view of
the fact that the barriers for diffusion of adatoms, dimers, and vacancies
are comparable, as we have just seen, we provide here a detailed discussion
of our MD/EAM simulations, with particular emphasis on prefactors which are
not available from static approaches, starting with the case of adatoms. (See
also Sec.\ 5, below.)

MD simulations were performed at several temperatures between 650 and 900 K.
The lower end of the range corresponds to the limit for accumulating proper
statistics, while the upper end corresponds to the onset of surface
disordering, i.e., spontaneous creation of adatom-vacancy pairs. At high
temperatures, ``exotic'' mechanisms, such as long exchanges involving several
atoms, are present but to a much lesser extent than the usual jump and
exchange mechanisms. In view of the much higher energy barriers associated
with these exotic processes, and the exponential behavior of the diffusion
coefficient (see Eq. \ref{arrhenius}), their contribution to mass transport
at low temperature will be negligible. These will therefore be ignored here,
since we are primarily interested in low-temperature growth.

\begin{figure}
\vspace*{-0.25in}
\epsfxsize=2.5in \epsfbox{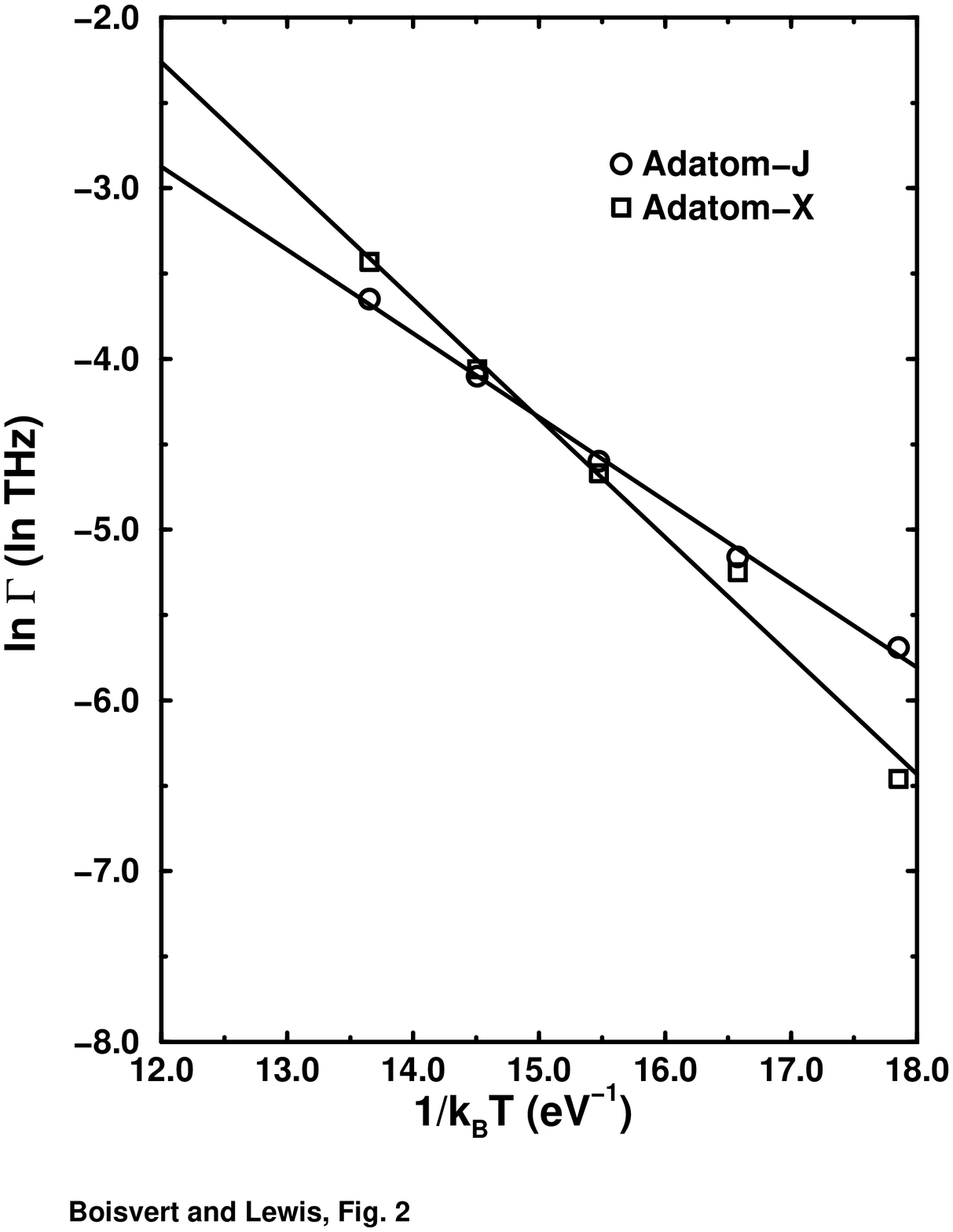}
\vspace{0in}
\caption{
Arrhenius plot of the frequency of jumps (J) and exchanges (X) for a Cu
adatom on Cu(100). The solid lines are fits to the MD data.
\label{Arr_mono}
}
\end{figure}

In Fig.\ \ref{Arr_mono}, we present Arrhenius plots of the frequency of jump
and exchange events for the adatom. The corresponding parameters ---
attempt-to-diffuse frequencies (prefactors) $\Gamma_0$ and energy barriers
$E_A$ --- are listed in Table \ref{para}. We find that the barrier for jump
diffusion is significantly smaller than that for exchanges, as was found also
in the MS calculations. In contrast, not available from MS, the prefactor for
exchange is found to be {\em much larger} --- by a factor of about 20 ---
than that for jumps, in qualitative agreement with the compensation law, eq.\
\ref{mnr}. This is an important result: Because the barrier for exchanges is
so much larger than that for jumps, the former would hardly be observable on
the MD timescale if it was not of compensation. In fact, from Fig.\
\ref{Arr_mono}, we see, as another consequence of compensation, that
diffusion crosses over from a regime where jumps predominate at temperatures
lower than $\sim$750 K to a regime where exchanges take over. If one thinks
in terms of mass transport, rather than frequencies, the crossover
temperature is even lower, $\sim$ 650 K, because the mean-square displacement
associated with an exchange event is twice as large as that for a jump (cf.\
Fig.\ \ref{diff_proc}). At low temperatures, evidently, compensation is not
strong enough to overcome the difference in barriers. For example, at 300 K,
using the present Arrhenius parameters, we would observe, on average, 150
jumps for a single exchange event.

\subsubsection{Dimers}

The barriers for dimer diffusion, we have found in the MS calculations, are
very similar to the corresponding ones for the adatom within EAM. It is thus
of interest to examine how prefactors compare in order to determine the
dominant contribution to mass transport.

\begin{figure}
\vspace*{-0.25in}
\epsfxsize=2.5in \epsfbox{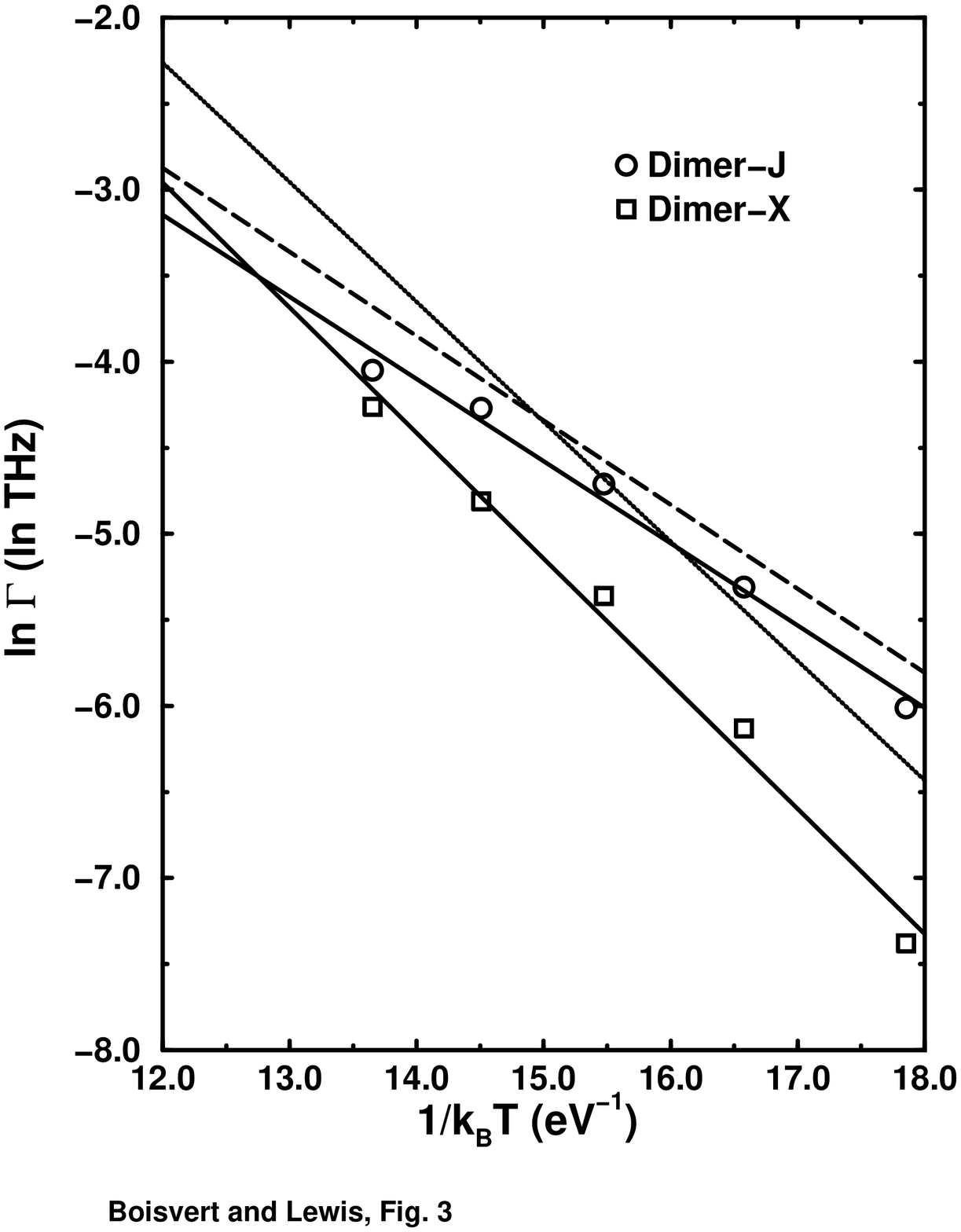}
\vspace{0in}
\caption{
Arrhenius plot of the frequency of jumps (J) and exchanges (X) for a Cu dimer
on Cu(100). The solid lines are fits to the MD data. The dashed and dotted
line correspond to adatom jumps and exchanges, respectively.
\label{Arr_dim}
}
\end{figure}

An Arrhenius plot of the frequency of jumps and exchanges is given in Fig.\
\ref{Arr_dim}; the corresponding parameters are listed in Table \ref{para}.
Again, here, exotic diffusion mechanisms can take place at high temperatures
(e.g., jumps involving the concerted motion of the atoms forming the dimer),
but they are present to a lesser extent than the two mechanisms depicted in
Fig.\ \ref{diff_proc} and can be neglected in the study of low-temperature
growth. The jump is the preferred mechanism for diffusion at low
temperatures, with a barrier about 0.25 eV smaller than that for exchanges.
The prefactors, however, show the opposite behavior, i.e., compensation again
is present. In this case, the crossover occurs at about 900 K, somewhat
higher than for adatoms.

It is quite remarkable that the jump and exchange diffusion barriers are the
same, within error, for adatoms and dimers. Likewise, the prefactors are
essentially equivalent: the observed differences, $\sim$50\%, are hardly
significant in that they could easily be absorbed in variations of the
exponential factor that could arise from small errors in the energy barriers.
Thus, for all practical purposes, the two species behave in a similar manner,
as can in fact be seen in Fig.\ \ref{Arr_dim}, and thus contribute equally to
mass transport within EAM.

\subsubsection{Vacancies}

\begin{figure}
\vspace*{-0.25in}
\epsfxsize=2.5in \epsfbox{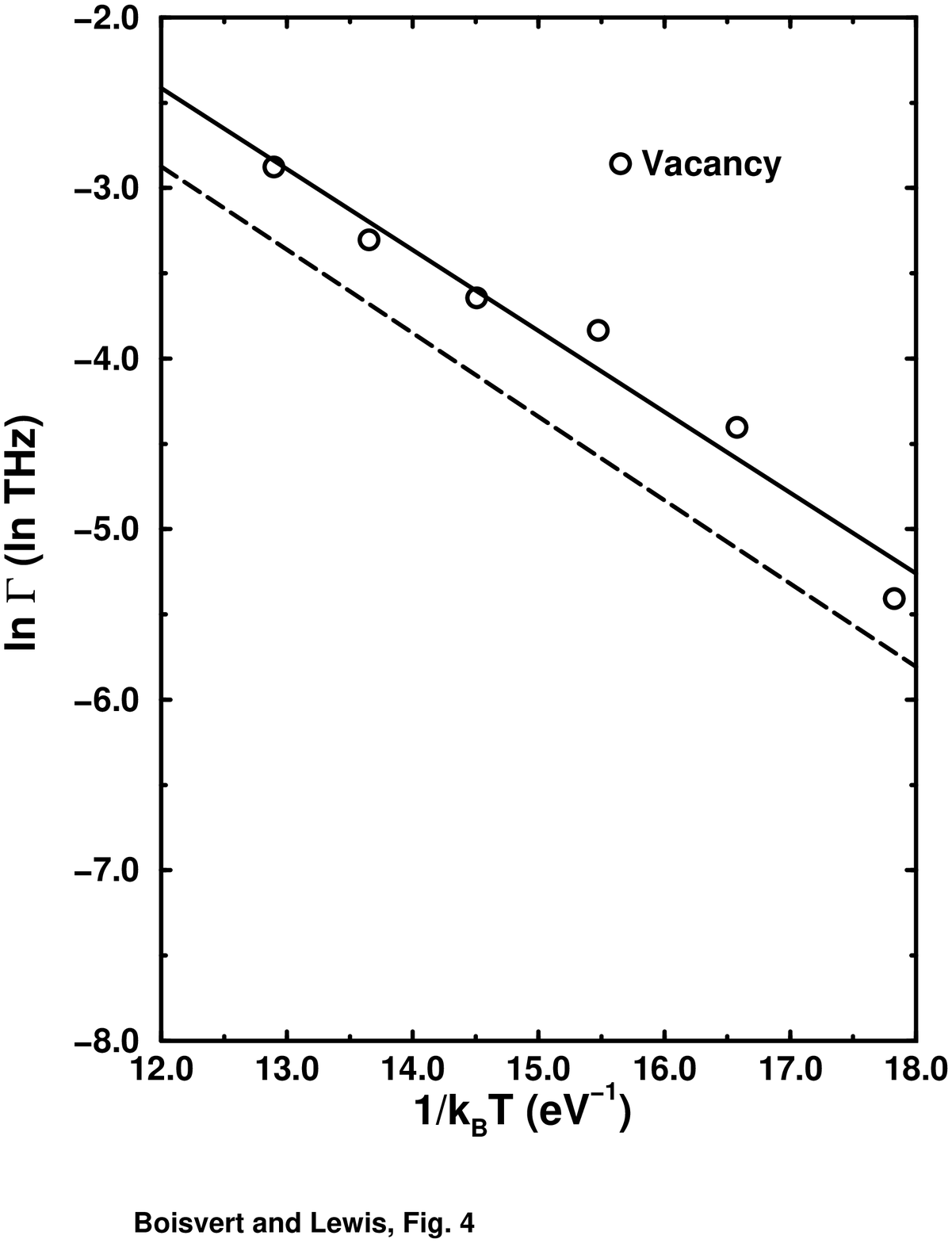}
\vspace{0in}
\caption{
Arrhenius plot of the frequency of jumps for vacancy diffusion on Cu(100).
The solid line is a fit to the MD data. The dashed line is the corresponding
frequency for adatom jumps.
\label{Arr_vac}
}
\end{figure}

Fig.\ \ref{Arr_vac} shows the Arrhenius frequency of jumps for the vacancy;
the corresponding parameters are listed in Table \ref{para}. No other
processes provide a significant contribution to diffusion though we have
observed, at high temperatures, some rare long-jump events. Clearly, the
vacancy and the adatom display similar behavior, with perhaps a slight edge
to the vacancy, both in terms of energy barriers (0.47 vs 0.49 eV) and
prefactors (27 vs 20 THz); the GGA predicts an even lower barrier for
vacancies. Thus, as far as mass transport is concerned, the two processes
contribute in essentially the same way during a single event.

\subsubsection{Steps}

\begin{figure}
\vspace*{-0.25in}
\epsfxsize=2.5in \epsfbox{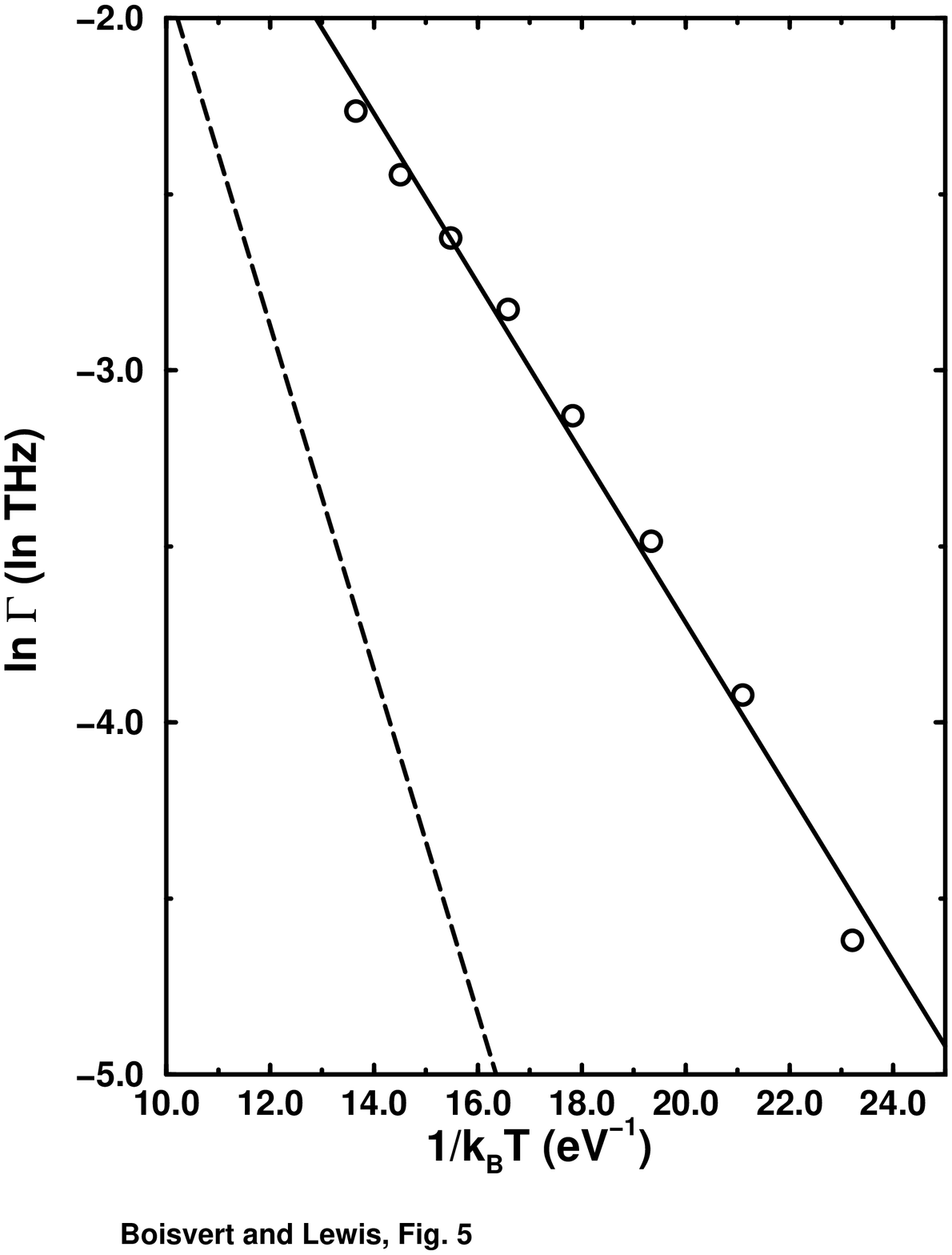}
\vspace{0in}
\caption{
Arrhenius plot of the frequency of events for diffusion of a Cu atom along a
step on Cu(100). The solid line is a fit to the MD data. The dashed line is
the frequency of the corresponding mechanism on the clean surface.
\label{Arr_step}
}
\end{figure}

Finally, in Fig.\ \ref{Arr_step}, we display the frequency of jumps for
diffusion along a $\langle$110$\rangle$ step on the (100) surface. The
frequency of jumps on the clean surface is also shown for comparison. The
Arrhenius parameters are given in Table \ref{para}. The barrier for diffusion
along the step is twice as small as that for jumps on an infinite, flat (100)
surface as was predicted from MS calculations. In spite of the fact that the
prefactor is roughly one order of magnitude smaller than on the terrace (3 vs
20 THz), diffusion along the step is much faster due to its relatively low
barrier. For instance, at 300 K, diffusion along a step is $\sim2000$ times
faster than on a terrace. Thus, it is certainly the case that an island, upon
the arrival of an adatom from the terrace, has time to rearrange itself into
its equilibrium shape before another adatom comes in. In other words, islands
remain close to equilibrium during growth.

\subsubsection{Final Remarks}

Before moving on to a discussion of our findings in the context of growth
experiments, a few remarks are in order. First, we find, in all cases
examined, the static activation energy to lie very close to the corresponding
barrier determined from detailed, extensive MD simulations, as can be seen
from Table \ref{para}. This indicates that, at least for the system under
consideration here, {\em accurate} energy barriers can be obtained from
purely static, first-principles calculations. This is at variance with the
results of Tully and coworkers,\cite{tully} who found, using a
``ghost-particle approach'' and a Lennard-Jones potential, the dynamical
barrier for the dimer to differ from the static one. Likewise, Evangelakis
and Papanicolaou,\cite{evangelakis} using a tight-binding description of the
interatomic potentials, found a dynamical barrier lower than the static one
for adatom exchanges on Cu(100). In the latter case, however, the statistics
are much poorer than ours; we found, in fact, that their diffusion data,
within the statistical uncertainties, can readily be accommodated by an
Arrhenius law with a barrier equal to the static value. Second, we find that,
given an energy barrier, the attempt-to-diffuse frequencies (prefactors) are
similar, regardless of the species undergoing diffusion. Third, although it
is difficult to draw firm conclusions from the data presented above, it seems
that, not surprisingly, the compensation law is valid not only for
adatoms,\cite{boisvert1} but also for dimers and vacancies. Based on our
second remark above, it would appear that the {\em same} set of Meyer-Neldel
parameters --- $X_{00}$ and $\Delta_0$ (cf.\ Eq.\ \ref{mnr} --- could
describe diffusion frequencies for the adatom, the dimer, and the vacancy;
more calculations are however required to assess this point in more detail.

\section{Discussion}\label{disc}

To our knowledge, there exists four different experimental determinations of
the diffusion barrier of a Cu adatom on
Cu(100).\cite{miguel,breeman1,ernst3,durr} As mentioned in the Introduction,
the values reported in these vary quite a bit, from 0.28 to 0.40 eV, and do
not agree with those calculated so far; diffusion on this surface, evidently,
is not well understood. We discuss here the results of our calculations in
the light of these experiments.

In the first experiment,\cite{miguel} the diffusion barrier was inferred from
a study of growth via step propagation. The diffusion coefficient, indeed,
can be related to the mean size of terraces; by measuring this quantity as a
function of temperature, between 318 and 415 K, and fitting to an Arrhenius
law, a barrier of 0.40 eV and a diffusion prefactor of $1.4\times10^{-4}$
cm$^2$/s, were obtained. The energy barrier is quite a bit smaller than the
one we obtained for adatom or dimer jumps, which dominate diffusion as we
have seen earlier, about 0.50 eV. The experimental prefactor corresponds to
an attempt-to-diffuse frequency of 0.8 THz, more than an order of magnitude
smaller than the EAM result. In view of the good agreement between EAM and
{\em ab initio} calculations for the barriers for jumps, we expect the
calculated attempt-to-diffuse prefactors to be correct within at most an
order of magnitude, thus in disagreement with the experimental value. In
fact, if we extrapolate the calculated diffusion coefficient to temperatures
in the range 318--415 K, we find agreement within a factor of two with
experiment, thus suggesting that indeed there is a possibility that both the
prefactor and the barrier in Ref.\ \onlinecite{miguel} are underestimated.
One possible explanation for the disagreement is that species other than
adatoms, such as dimers, might be present experimentally, in view of the
relatively high temperature, and contribute to diffusion.

Using low-energy ion scattering (LEIS), second, Breeman and
Boerma\cite{breeman1} obtained a diffusion barrier of $0.39 \pm 0.06$ eV,
{\em assuming} a prefactor of 10 THz --- quite a bit larger than the value of
0.8 THz estimated from step propagation measurements (see above). In these
experiments, adatoms are created by the ion-beam irradiation of a surface.
Their concentration can be estimated from the LEIS yield, which changes as a
function of temperature because of diffusion towards --- and incorporation
into --- the steps between terraces (sometimes referred to as ``annealing'').
An abrupt change in the LEIS yield signals the onset of adatom mobility;
given the timescale of the experiment and the length of the terraces, it is
then possible to determine the diffusion coefficient. Assuming a value for
the the prefactor, finally the diffusion barrier can be extracted. It should
be stressed that these measurements are carried out at a single temperature;
this is the reason the prefactor must be assumed in order to determine the
activation energy, leading to possibly large errors.

In addition to adatoms, however, surface vacancies can also be created during
irradiation. It is not clear what their effect is on annealing. Our
calculations indicate that they have a diffusion barrier lower than adatoms.
Thus they could, for example, recombine with neighboring, immobile, adatoms.
Experiment, therefore, would measure the onset of mobility of vacancies
rather than adatoms. This question has been discussed in Ref.\
\onlinecite{breeman3}, where it is argued, on the basis of an empirical
model, that vacancies start diffusing at about 120 K and are all annealed
(into steps) by the time temperature reaches the adatom mobility edge, about
140 K; i.e., vacancies would {\em not} affect diffusion. However, our {\em
ab-initio} barrier for vacancy diffusion is in good agreement with the
activation energy determined from LEIS --- 0.42 vs 0.39 eV --- and we must
therefore conclude that the diffusion of vacancies remains a possible
explanation for the observed onset of mobility. We note that in the case of
Ag/Ag(100)\cite{langelaar} (see also Sec.\ \ref{ab_res_ad}), also using LEIS,
theory and experiment are in excellent agreement. It would be interesting to
determine the diffusion barrier of vacancies in this case and see how it
compares to adatom diffusion. If our interpretation is correct, vacancies on
Ag(100) should not be more mobile than adatoms on the same surface.

Finally, in view of the relatively small size of terraces in this last
experiment, 8.5 atomic spacings, it is not clear that this geometry can
effectively be used to determine diffusion barriers appropriate to infinitely
wide terraces: In the case of Ir/Ir(111), for instance, it was noted (using
FIM) that no adatoms are ever found in a region of width 3 nearest-neighbor
distances from steps,\cite{wang} likely the consequence of a lower diffusion
barrier in the vicinity of steps. Such an effect could bias experimental
estimates of the barriers in cases where the depletion zone is a large
fraction of the diffusion length, possibly the case in the above LEIS
measurements.

In the last two experiments of interest,\cite{ernst3,durr} the separation of
islands was measured and related to energy barriers through rate equations,
assuming adatoms are the only mobile species (i.e., all larger clusters
remain immobile and unable to dissociate.) While the published results are
different, $0.28 \pm 0.06$ eV (Ref.\ \onlinecite{ernst3}) and $0.36 \pm 0.03$
eV (Ref.\ \onlinecite{durr}), the data of Ref.\ \onlinecite{ernst3} was
recently reinterpreted,\cite{ernst4} leading to a value of about 0.40 eV, in
line with other experiments.

{\em Grosso modo}, from the above experiments, the diffusion barrier can be
taken as $0.40 \pm 0.05$ eV. This is reasonably close to our GGA value (for
adatom jumps) of $0.52 \pm 0.05$ eV, but the deviation is large enough to
warrant closer examination. There are evidently two possibilities. First, the
theoretical value may be in error, either because of model limitations (e.g.,
size), or because of a poor description of the exchange-correlation energy.
We have at present no way of assessing these further. Second, it is possible
that the assumptions underlying the interpretation of experimental data may
not be fully justified. We have mentioned already that vacancies or limited
terrace size could possibly play a role in the interpretation of LEIS
measurements. In what follows, we examine more closely the assumptions behind
rate equations.

In a rate-equation analysis, the diffusion coefficient depends on the island
separation through a power law. If only adatoms are mobile, the exponent is
6. In a plot of the logarithm of island separation versus inverse
temperature, the slope is simply the diffusion barrier for adatoms divided by
this exponent. If dimers (and only dimers) can dissociate, then the exponent
is 4 and the slope is now related not only to the adatom barrier but to the
sum of adatom barrier and binding energy per atom of the dimer. Thus,
clearly, detailed knowledge of the surface kinetics, as well as highly
accurate data, are essential for extracting meaningful numbers from such
measurements. Our calculations indicate that the diffusion barriers for
adatoms and dimers are very close to one another and suggest, therefore, that
the assumptions underlying the rate-equation analysis might not be valid.

A first assumption concerns the stability, against diffusion and
dissociation, of small clusters, which determines the exact form of the
island-separation--diffusion-coefficient scaling relation. Experiment
suggests that, at low flux and low enough temperatures --- below 223 K ---
only adatoms are mobile.\cite{zuo} Above 223 K, dimers and trimers can
dissociate, and thus change the scaling relation. According to our results,
however, as discussed in Sec.\ \ref{ab_res_dim}, dimers should be mobile
before they can dissociate. At 223 K, indeed, we find the rate of jump for
dimers (i.e., non dissociated) to be approximately 10\% that for adatoms
(assuming similar prefactors). This, of course, affects the scaling relation
and, therefore, the value of the barrier that can be inferred from the
experimental data. In a recent Monte Carlo study of nucleation on
Pt(111),\cite{bott} it was found that the island density remains unaltered in
presence of dimer diffusion, as long as the barrier for the latter is {\em at
least} 0.09 eV higher than the barrier for adatom diffusion. The difference
between the two barriers here is 0.05 eV, thus suggesting that dimer mobility
{\em cannot} be neglected.

Interestingly, the island separation at this same temperature, 223 K, can be
reproduced by Monte Carlo simulations assuming that islands are square and
that adatoms only are mobile.\cite{bartelt2} The jump frequency required to
obtain satisfactory agreement with experiment is found to be 450 s$^{-1}$.
Assuming a prefactor of 20 THz, as obtained in the present work, this
translates into an energy barrier of 0.47 eV, now within the error bar of our
theoretical prediction. This is strong indication that an imperfect scaling
relation in the rate equations can lead to significant errors in experimental
diffusion barriers.

When temperature is higher than 223 K, dimers are found to dissociate. Using
rate equations, and assuming the smallest stable (``critical'') island to be
the tetramer, D{\"u}rr and coworkers\cite{durr} estimated a binding energy of
0.08 eV for the dimer. This is much less than the value we find --- 0.35 eV
from EAM and 0.22 eV from first principles (cf.\ Table \ref{dim}). However,
these data were recently reinterpreted by Bartelt and
coworkers.\cite{bartelt1} They found that the change of the scaling relation
occurring at 223 K is due to a gradual transition in critical island size,
related to the onset of dimer dissociation, and not to a sharp transition
from the adatom to the tetramer as assumed in Ref.\ \onlinecite{durr}. In
this way, a dimer binding energy of 0.20$-$0.23 eV is obtained, in excellent
agreement with the present calculations. It should be said, however, that the
latter value was obtained assuming an energy barrier of 0.40 eV for the
adatom, rather than 0.52 eV from the present theory.

To conclude on this point, it appears that dimer mobility has to be taken
into account in order to describe correctly low-temperature growth on
Cu(100). (We have not explored, because of computer limitations, the
possibility that trimers also contribute, but this should not be completely
ruled out.) This results in a very complicated scaling relation and therefore
potentially significant errors in estimates of the energy barriers for
diffusion.

We now discuss the shape of islands. We have found, from our MD/EAM
simulations, that the barrier for diffusion along steps is much smaller than
that for diffusion on flat (100) surfaces --- 0.26 vs 0.50 eV. (We expect, in
view of the agreement for other barriers, that {\em ab initio} calculations
would lead to equivalent results.) Thus we predict that the shape of islands
will remain close to equilibrium, i.e., square, during growth as indeed is
observed experimentally.\cite{durr} There exists, to our knowledge, only two
experimental reports of this barrier, and they disagree sharply: In Ref.\
\onlinecite{poensgen}, a barrier of approximately 0.1 eV is given, consistent
with the observed island shape. The other, Ref.\ \onlinecite{girard1}, in
contrast, reports a very high value of 0.45 eV, comparable to the barrier for
diffusion on (100) terraces, and very likely too high to yield the correct
island shape. Clearly, more measurements are needed to resolve this point.

\section{Concluding Remarks}\label{rem}

We have presented a detailed study of the diffusion of adatoms, dimers, and
vacancies on Cu(100), using both {\em ab initio} static relaxation methods
and semi-empirical simulations. Our results are discussed in the context of
recent submonolayer growth experiments. We find that the GGA offers a much
better description of the energetics of diffusion than the LDA, while the EAM
yields generally satisfactory results, in addition to providing information
on attempt-to-diffuse frequencies (prefactors). Vacancy diffusion is found to
be the most favorable mechanism for mass transport, but is not necessary
dominant, as it depends on details of the experiments. The value we obtain
for the energy barrier for adatom diffusion is slightly larger than the
available experimental numbers. However, we have demonstrated that the
complexity of the scaling relations obtained from rate equations (and used to
interpret the experimental measurements), arising for instance from small
cluster mobility, could easily explain this discrepancy: Indeed, we have
found that, at low temperatures, dimers are mobile, though to a lesser extent
than adatoms. Dimers are also found to diffuse more readily than they
dissociate. The preferred diffusion mechanism is the jump, for both the
adatom and the dimer, i.e., exchange processes do not seem to be an important
route for diffusion on this surface at low temperatures. Finally, our MD
study of diffusion of adatoms, vacancies, and dimers revealed no clear
dependence of the prefactors on the diffusing species and, in all cases, the
static barrier was found to approximate well the dynamical barrier. From the
present study, we conclude that a combination of highly-accurate {\em ab
initio} static calculations and semi-empirical MD simulations provides a good
basis for determining diffusion processes relevant to growth.

\acknowledgements

We are grateful to Martin Fuchs for help with the pseudopotential generation
and to Normand Mousseau, Risto Nieminen, Christian Ratsch, Ari Seitsonen,
Matthias Scheffler, and Byung Deok Yu for stimulating discussions. This work
was supported by grants from the Natural Sciences and Engineering Research
Council (NSERC) of Canada and the ``Fonds pour la formation de chercheurs et
l'aide {\`a} la recherche'' (FCAR) of the Province of Qu{\'e}bec. One of us
(G.B.) is thankful to NSERC and FCAR for financial support. We are grateful
to the ``Services informatiques de l'Universit{\'e} de Montr{\'e}al'' for
generous allocations of computer resources. Part of the work reported here
has been performed on the IBM/SP-2 from CACPUS (``Centre d'applications du
calcul parall{\`e}le de l'Universit{\'e} de Sherbrooke'').


\newpage

\onecolumn
\widetext

\begin{center}
\begin{table}
\caption{
{\em Ab-initio} results for bulk Cu and and the clean (100) surface; FP-LMTO
and PP-PW refer to the all-electron and pseudopotential calculations from the
present work. AE is for other all-electron calculations and PP for other
pseudopotential calculations. LDA and GGA specify the level of approximation
used for the exchange-correlation energy.
}
\label{tests}
\begin{tabular}{lccccc}
            & Lattice constant & Surface energy & \multicolumn{2}{c}{Surface relaxation} & Work function \\
            &     $a$          &   $\sigma$     & $\Delta d_{12}$  &  $\Delta d_{23}$    &      W       \\
            &    (\AA)          &   (J/m$^2$)      &   ($\% d_{bulk}$)  &     ($\% d_{bulk}$)   &      (eV)      \\ \tableline
FP-LMTO-LDA &    3.50          &    1.85        &      -3.0        &        -            &     4.87     \\
PP-PW-LDA   &    3.57          &    1.91        &      -3.5        &       0.0           &     4.86     \\
PP-PW-GGA   &    3.68          &    1.42        &      -4.5        &      -0.4           &     4.42     \\
AE-LDA    &3.52\cite{khein,korhonen}, 3.55\cite{polatoglou}, 3.56\cite{lu}, 3.58\cite{kraft}, 3.61\cite{lu} & - & - &   -      &  -       \\
AE-GGA    & 3.62\cite{khein}       &     -          &         -        &        -            &       -  \\
PP-LDA  &3.62\cite{rodach,kang,cheli}, 3.61\cite{jeong} &1.71\cite{rodach}&-3.02\cite{rodach}&0.08\cite{rodach}&4.95\cite{rodach}\\
expt.  &3.60\cite{touloukian}&2.02\cite{richter}&-1.2\cite{lind}&0.9\cite{lind}&4.59\cite{gartland}, 4.83\cite{haas}, 4.45\cite{peralta}\\
\end{tabular}
\end{table}
\end{center}

\begin{center}
\begin{table}
\caption{
Diffusion barriers and prefactors for diffusion on Cu(100). The {\em ab
initio} values are obtained using the GGA with a $4$-layer, ($3 \times 3$)
cell. J and X are for jumps and exchanges, respectively. $E_A^0$ is the
zero-temperature (static) value of the energy barrier while $E_A$ and
$\Gamma_0$ are determined from an Arrhenius fit to the MD data.
}
\label{para}
\begin{tabular}{|l|ccc|c|}
           &    \multicolumn{3}{c|}{EAM}                      & {\em ab initio}  \\ \tableline
           & $\Gamma_0$ (THz) & $E_A$ (eV)      & $E_A^0$ (eV) &  $E_A^0$ (eV)      \\ \tableline
Adatom-J   & 20(e$^{\pm0.2}$) & 0.49 $\pm$ 0.01 & 0.50       &  0.52 $\pm$ 0.05 \\
Adatom-X   &437(e$^{\pm0.7}$) & 0.70 $\pm$ 0.04 & 0.73       &  0.96 $\pm$ 0.10 \\
Dimer-J    & 13(e$^{\pm0.5}$) & 0.48 $\pm$ 0.03 & 0.49       &  0.57 $\pm$ 0.06 \\
Dimer-X    &320(e$^{\pm0.8}$) & 0.73 $\pm$ 0.05 & 0.74       &  0.79 $\pm$ 0.15 \\
Vacancy    & 27(e$^{\pm0.7}$) & 0.47 $\pm$ 0.05 & 0.47       &  0.42 $\pm$ 0.08 \\
Along step & 3.0(e$^{\pm0.2}$)& 0.24 $\pm$ 0.02 & 0.26       & -                \\
\end{tabular}
\end{table}
\end{center}

\begin{center}
\begin{table}
\caption{
Static properties of the dimer. The {\em ab initio} values are obtained using
the GGA with a $4$-layer, ($3 \times 3$) cell. See the text for a definition
of the exact and approximate forms of the dissociation energy.
}
\label{dim}
\begin{tabular}{lcc}
                               &   EAM    & {\em ab initio}  \\
                               &   (eV)   &   (eV)           \\ \tableline
Binding energy                 & 0.35     &   0.22 $\pm$ 0.03\\
Dissociation energy (exact)    & 0.81     &   -              \\
Dissociation energy (approx.)  & 0.85     &   0.74 $\pm$ 0.07\\
Metastable vs equilibrium      & 0.29     &   0.35 $\pm$ 0.04\\
\end{tabular}
\end{table}
\end{center}

\begin{center}
\begin{table}
\caption{
Diffusion barrier for the adatom (in eV) from first principles, as discussed
in the text. The superscript ``u'' is for an unrelaxed substrate while ``2''
refers to the case where the top two layers of the substrate were relaxed;
in all others, only the top layer relaxed only. ``BZ'' refers to a denser
{\bf k}-point grid for Brillouin zone integration.
}
\label{adatom}
\begin{tabular}{|l|c|cc|cc|}
  System      &  FP-LMTO &              \multicolumn{4}{c|}{PP-PW}                   \\
              &  LDA     &   \multicolumn{2}{c}{LDA}       &\multicolumn{2}{c|}{GGA} \\ \tableline
              &   jump   &   jump         & exchange       &   jump   & exchange    \\
(2 x 2) cell: &          &                &                &          &             \\
 ~ 3 layers   & -        & 0.75$^u$, 0.74 & 1.23$^u$, 1.18 &      -   &     -       \\
 ~ 4 layers   & -        & 0.66, 0.65$^2$, 0.68$^{BZ}$ & 1.05, 1.04$^2$, 1.08$^{BZ}$ &   0.51   &  0.85       \\
 ~ 5 layers   & 0.69$^u$ &   0.69         &   1.04         &   0.55   &  0.82       \\
 ~ 6 layers   & -        &   0.65         &   1.18         &    -     &    -        \\
 ~ 7 layers   & 0.66$^u$ &   0.69         &   1.13         &    -     &     -       \\
 ~ 9 layers   & 0.68$^u$ &   -            &   -            &     -    &   -         \\
(3 x 3) cell: &          &                &                &          &             \\
 ~ 3 layers   & -        &   0.75         &   1.03         &     -    &        -    \\
 ~ 4 layers   & -        &   -            &    -           &   0.52   &  0.96       \\
 ~ 5 layers   & 0.65$^u$ &   -            &    -           &     -    &     -       \\
\end{tabular}
\end{table}
\end{center}


\input{epsf.tex}

\vfill

\twocolumn
\narrowtext

\vspace{0.23in}

\end{document}